\documentclass[12pt,preprint]{aastex}
\begin{document}

\title{POLAR FIELD PUZZLE: SOLUTIONS FROM FLUX-TRANSPORT DYNAMO 
AND SURFACE TRANSPORT MODELS}
\author{MAUSUMI DIKPATI}
\affil{High Altitude Observatory, National Center for Atmospheric
Research \footnote{The National Center
for Atmospheric Research is sponsored by the
National Science Foundation. },
3080 Center Green, Boulder, Colorado 80301; dikpati@ucar.edu}

\begin{abstract}
Polar fields in solar cycle 23 were about 50\% weaker than
those in cycle 22. The only theoretical models which have addressed 
this puzzle are surface transport models and flux-transport dynamo
models. Comparing polar fields obtained from numerical simulations 
using surface flux transport models and flux-transport dynamo models, 
we show that both classes of models can explain the polar field 
features within the scope of the physics included in the respective 
models. In both models, how polar fields change as a result of 
changes in meridional circulation depends on the details of 
meridional circulation profile used. Using physical reasoning and 
schematics as well as numerical solutions from a flux-transport 
dynamo model, we demonstrate that polar fields are determined 
mostly by the strength of surface poloidal source provided by the decay 
of tilted, bipolar active regions. Profile of meridional flow 
with latitude and its changes with time have much less effect in 
flux-transport dynamo models than in surface transport models.

\end{abstract}
\keywords{Sun: magnetic field, Sun: activity, Sun: dynamo}

\section{Introduction}

Observations from various instruments indicate that the polar fields
in cycle 23 were so weak (see figure 1 of \citet{wrs2009}, see also 
figure 2 of \citet{ahph2002} and figure 2 of \citet{jbg2010}) that it took 
a relatively long time to reverse the magnetic polarity of the Sun's 
North and South poles. Even after the Sun's polarity reversal the 
build-up of the polar fields was slow \citep{ddgaw2004}, resulting 
in unusually weak polar fields at the end of cycle 23 compared to 
those in cycles 21 and 22. In order to understand the cause of such 
a weak polar field at the end of cycle 23 and its consequences for the 
solar-terrestrial environment, many scientists have extensively employed 
simulations from flux-transport dynamo models \citep{dc99,ddg2008,dgdu2010} 
and surface transport models \citep{wls2005,sdt2002,sl2008}. 

In both these classes of models the polar fields originate
from the decay of tilted, bipolar active regions, namely the so-called
Babcock-Leighton mechanism \citep{babcock1959, leighton1964}. The ingredients
(advection and diffusion) that directly influence the latitudinal transport
of the radial fields in the two classes of models are the same. However,
one of the differences in the evolutionary patterns of the polar fields 
is that, in flux-transport dynamo models, poloidal fields produced 
at the surface by the Babcock-Leighton effect are axisymmetric (i.e. in 
the $r-\theta$ plane) and they evolve due to advection and diffusion by 
the action of both the latitudinal and radial components of the meridional 
flow as well as by the action of a depth-dependent turbulent diffusivity. 
By contrast, surface transport models include a more realistic longitude 
dependence of the Babcock-Leighton effect in the generation of polar fields, 
and the radial component of the fields generated evolve at the surface by 
the action of latitudinal component of the meridional flow and a constant 
surface diffusivity.

The purpose of this paper is to demonstrate that flux-transport dynamo
models and surface transport models, despite some differences in the
ingredients, produce remarkably similar response in the polar
fields' patterns to the changes in the meridional flow-speed when 
the same latitudinal profile for the poleward surface flow is used.
Given that understanding of the polar fields' behaviour is
a key to understanding the recycling of magnetic flux for the operation
of a dynamo, and hence the properties of future solar cycles, we also 
examine in this paper whether there are any inconsistencies or 
contradictions between these two classes of models when applied to
the Sun. The polar field puzzle of cycle 23 has become a 
far more important issue now that the start of cycle 24 has been sluggish.

We illustrate in simple numerical terms just how sensitive the amplitude 
of the polar field on the Sun and in models is to changes in the strength 
of the source of its field, namely the emergence and decay of active 
regions. We know that the amplitude of cycle 23 was 20\% weaker than 
that of cycle 22. If at the end of a given 
cycle the polar field has one unit, in the next cycle it takes two units 
of polar fields coming from the surface poloidal source to reverse the 
remnant polar field and have the new polar fields reach minus one unit. 
But if the surface poloidal source is 20\% smaller in the new cycle than 
the previous one, as cycle 23 was compared to cycle 22, there are only 
1.6 units of new, negative, polar fields available that can be used to 
reverse the old, positive, polar field and establish the new, negative, 
polar field. This means the new, negative, polar field will be only 
0.6 units, or 40\% smaller in amplitude than that of the previous cycle. 
The percentage decline in polar field is, therefore, roughly twice the 
decline in surface source. The observed decline in polar field between 
cycles 22 and 23 was about 50\%. Thus at the outset most of this change 
can be explained by the change in surface source, whereas other effects
are needed to account for the additional drop of 10\%. 

For both the surface transport and flux-transport dynamo models polar 
fields are the follower of a cycle by virtue of their origin, namely 
the decay of tilted, bipolar active regions, hence in a crude analysis 
as described above, the 40\% reduction in polar fields at the end of 
cycle 23 might be true for all such models. However, a potential
weakness of the reasoning just given above is the assumption of a close 
relation between the strength of a cycle, as measured by the Wolf sunspot 
number or sunspot area, and the strength of the surface poloidal source
arising from the decay of active regions. That is undoubtedly
an oversimplification. It is well known that there is a strong 
correlation between Wolf number and sunspot area; the latter is what
has usually been used to drive flux-transport dynamo models and
to compare with model output. As seen for example in figure 1 of
\citet{ddg2006} there is a good correlation between sunspot area and
average surface magnetic flux, provided both data sets are averaged
over at least six solar rotations. So presuming a good correlation 
between Wolf number and the surface source could be plausible, but
recent surface transport models \citep{wrs2009} do a better job of 
estimating the surface source from more detailed distributions of 
active region sources. 

\section{Role of meridional circulation in polar fields' evolution}

The structure and strength of the meridional circulation influence
the strength of the Sun's magnetic fields to some extent, but 
meridional circulation cannot be the most important factor determining the
field strength in flux-transport dynamo models. In flux-transport
dynamos (and most other dynamos applied to the sun) the spot-producing
toroidal fields are generated by the Sun's differential rotation and the
poloidal fields are produced by the action of the so-called 
$\alpha$-effect. The $\alpha$-effect is modeled in different dynamo
models in different ways. In Babcock-Leighton type flux-transport dynamo
models, such an $\alpha$-effect arises from the decay of tilted, bipolar
active regions that emerge to the surface from below. Thus the two
sources of magnetic fields, the differential rotation and a 
combination of $\alpha$-effects arising from helical turbulence as
well as from the decay of active regions, are primarily responsible for
determining the amplitudes of toroidal and poloidal fields, and hence
the polar fields.

The primary role of meridional circulation in flux-transport dynamo
models is the advective transport of magnetic fields, and hence the
structure and strength of the flow are crucial in the dynamo models for
determining the timings, namely the duration of a cycle, its rise and
fall pattern and the timing of the reversals of the Sun's polar fields
\citep{dc99,dikpati2004,dgdu2010}. A transport process
like meridional circulation in a flux-transport dynamo redistributes the
dynamo-generated magnetic flux, a relatively minor effect in creating an
increase or decrease of magnetic flux. This is illustrated in Table 1 of
\citet{dc99} in which it is seen that, if the
meridional circulation speed is doubled, the peak polar field changes
by only $3\%$, whereas for the same meridional circulation, increasing
the surface poloidal source by a factor 2.5 doubles the peak polar
field, and decreasing the surface source by $75\%$ decreases the polar
field peak by $89\%$.

The question we address is how flux-transport dynamo models and surface 
transport models respond to the changes in the surface poleward flow speed. 
Detailed calculations show that a surface transport model produces a 
weaker polar field \citep{wls2005, sl2008}, while a flux transport dynamo 
gives a stronger polar field when the poleward surface flow-speed is 
increased \citep{dc99,ddg2008}. Are these opposite results in conflict?
We address, first with qualitative reasoning and schematics and then in 
the next section with numerical solution from a flux-transport dynamo 
model, whether this is a true conflict, and if so, what is the physical 
origin of this conflict. 

Figure 1 illustrates three scenarios: (a) and (b) for surface flux transport
models and (c) for a flux transport dynamo. The primary difference
between figures 1(a) and (b) is the latitude where the surface poleward
flow is maximum. Figure 1(a) illustrates what happens to surface flux 
in the declining phase of a cycle
when the meridional flow peaks at $6^{\circ}$ as in the model by 
\citet{wls2005}; Figure 1(b) illustrates what happens for a meridional 
flow that peaks at $37.5^{\circ}$ \citep{sdt2002}.
Figure 1(c) shows what happens to polar fields when a meridional
circulation profile in the pole-to-equator meridional plane peaks at
mid-latitudes $\sim 40^{\circ}$, as has been used in flux-transport
dynamo models by \citet{dc99} and \citet{ddg2008}.

\clearpage
\begin{figure}[hbt]
\epsscale{0.65}
\plotone{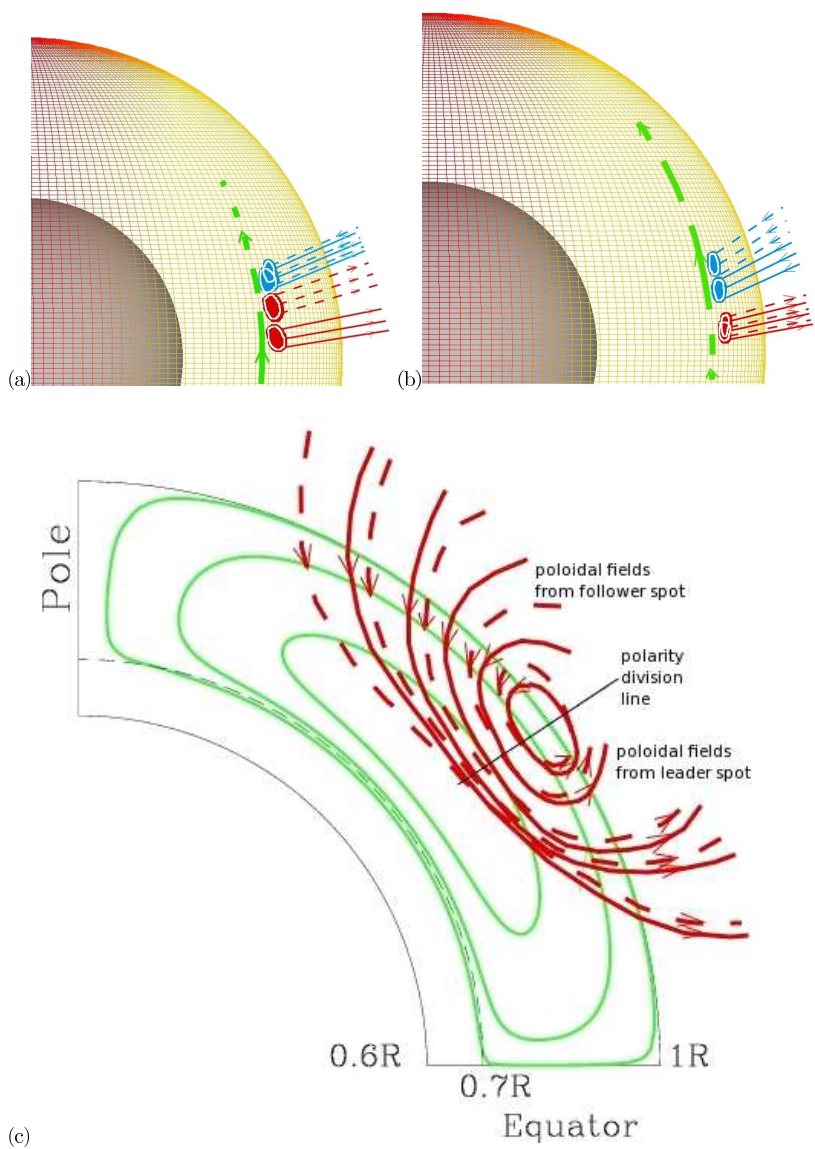}
\caption{Surface radial field transport mechanisms respectively
in a surface flux-transport model with a poleward flow peaking at
$6^{\circ}$ (frame a) as in the model of Wang et al. (2005), the 
surface transport model with a poleward flow peaking at $37.5^{\circ}$ 
(frame b) as in the model of Schrijver et al. (2002), and in a 
flux-transport dynamo model with a flow peaking at mid-latitudes 
(frame c) as in the model of Dikpati et al. (2008). In frames
(a) and (b), red and blue patches on the surface represent bipolar active
regions. Red and blue continuous lines represent radial fields from a
bipole, and dashed lines represent the new locations of radial fields
after the meridional flow is increased. For example, by doubling
flow speed, a larger increase in transport-rate is obtained
respectively on equatorward and poleward side of the bipoles in frames
(a) and (b); thus fields from leader polarity drift
closer to fields from follower polarity in frame (a) and
further apart in frame (b). Frame (c) describes a very similar
situation as in frame (b), but in terms of poloidal fields in 
$r-\theta$ plane. Solid contours represent poloidal fields produced
from bipolar active regions and dashed contours represent that for
an increased flow. Polarity division line of large-scale poloidal
fields from a flux-transport dynamo model is shown by a dark line.
}
\label{schematic}
\end{figure}
\clearpage

We can understand from Figure 1(a) that, when the flow is maximum at low
($6^{\circ}$) latitudes, an increase in poleward flow speed leads to a larger
rate of transport on the equatorward side of the leader polarity, and hence
a faster poleward transport of the leader polarity compared to the follower
polarity. Thus the leader polarity gets closer to the follower polarity to
create enhanced annihilation between them. This reduces the fields coming
from the follower polarity that reach the pole and thus produce weaker polar
fields. This is what has been obtained by \citet{wls2005} and \citet{sl2008}
in simulations using surface transport models.
On the other hand, when the flow speed is maximum at mid-latitudes
(Figure 1(b)), an increase in flow speed leads to a larger rate of transport 
on the poleward side, causing the follower and the leader polarities to 
separate more from each other. Thus there is less annihilation between them. 
Consequently the polar field should increase. \citet{sdt2002} explained, using
a flow peaking at $37.5^{\circ}$,
that the poleward meridional flow is so effective in maintaining the high 
latitude field that this flow would have to practically disappear to get a 
significant decay of the polar flux. We can thus infer that, quite consistently 
a decrease in flow leads to a decrease in polar flux, and increase in flow 
to an increase in polar flux when the flow peaks at $37.5^{\circ}$ latitude. 

It follows from the reasoning above that
the increase in polar field amplitude with the increase of meridional flow 
speed in flux-transport dynamo models \citet{dc99} and \citet{ddg2008}
and surface transport model of \citet{sdt2002}, and the decrease
in polar field with the increase in meridional flow-speed in the surface
transport models of \citet{wls2005} and \citet{sl2008}, are 
not in conflict. Rather, it is the choice of meridional flow profile --
peaking at low latitudes or at mid-to-high latitudes -- that leads to the 
opposite results which are physically consistent. We recognize that 
cross-equatorial flux cancellation may also play a role in producing the 
results of \citet{sl2008}.

There now exist detailed measurements of meridional flow speed by several 
methods for all of cycle 23, \citep{ulrich2010, ba2010, hr2010} so detailed 
comparisons can be made. This has been done most extensively in 
\citet{ulrich2010}. All measurements show significant variations from 
year to year within cycle 23, some of which appear to correlate with 
variations in rotation, in particular the torsional oscillations, as seen 
in the helioseismic results of \citet{ba2010}.  In terms of meridional 
flow averaged over cycle 23 (see figure 1 of \citet{dgu2010} and figure 10 
of \citet{ulrich2010}) it is clear that surface Doppler \citep{ulrich2010}
and helioseismic \citep{ba2010} results agree closely with each other, 
while the magnetic feature tracking results \citep{hr2010} are distinctly 
different. In particular, surface Doppler and helioseismic measures have 
an average peak at $25^{\circ}$ and the peak shifting with time between 
15 and 40 degrees, while magnetic feature tracking flow speed peaks 
closer to 50 degrees and has a distinctly lower peak value.

Early surface transport models \citep{dbs1984,wns1989} were the first 
to use a latitudinal meridional flow profile peaking at low latitudes, 
in particular, 6 degrees, even before observations could tell us where 
the flow peaks. Such a low-latitude peak gave the best fit of model-derived 
surface features with observations of surface magnetic fields. In future
the much more extensive Doppler based observations of meridional flow 
can be used as input to both surface transport models and flux-transport 
dynamo models, capturing the low latitude peak now observed. By contrast, 
the magnetic feature-tracking speed should be compared with the output 
of such models in order to estimate the surface magnetic diffusivity.

There is a tracking speed that might also be useful as input to dynamo
and surface transport models. That is the tracking speed that comes from
the Doppler signal of supergranules \citep{sksb2008}. It has been
used to measure differential rotation and meridional circulation for
solar cycle 23. It will be useful to compare the details of flow
profiles obtained by this method with those obtained by more global
surface Doppler as well as helioseismic measures.

\section{Model calculations}

\subsection{Polar field simulations from a self-saturated flux-transport 
dynamo driven by Babcock-Leighton $\alpha$-effect only}

In most flux-transport dynamo models, starting from \citet{dc99}, the 
poleward surface flow has
generally been taken to be a maximum in midlatitude. So according to the
qualitative reasoning given in Figure 1, the fields from the follower
polarity separate out from the leader polarity at a faster rate (see
Figure 1(c)) when the flow speed is increased. Thus there is less
annihilation among them; consequently the polar field increases.
To test this with flux-transport dynamo simulations, we must take
account of the fact that there are additional physical processes at
work in the model that will change the poloidal and toroidal fields
in other parts of the dynamo domain in response to a change in
meridional flow. For example, if the flow toward the poles at the top
speeds up, the return flow near the bottom will also speed up. This
bottom flow then moves the poloidal field there faster toward the
equator, leaving less time to induce toroidal field at a particular
latitude. This reduction in bottom toroidal field in turn leads to a
reduced surface poloidal source. Which wins in determining whether
the polar fields increase or decrease? In Figure 2a we show results from
new simulations with our flux-transport dynamo model using ingredients
very similar to those used in \citet{dc99} for a pure
Babcock-Leighton dynamo (no bottom $\alpha$-effect) that answers this
question. Figure 2a displays measures of the surface polar field (blue
diamonds) and the tachocline toroidal field (black triangles) for a
sequence of independent simulations of the self-excited dynamo for
different peak meridional flow speeds (all peaking at the same
midlatitude). Each simulation is run for about 5 cycles
to reach a nonlinear equilibrium in which
each successive cycle has the same period and amplitude.

\begin{figure}[hbt]
\epsscale{0.9}
\plottwo{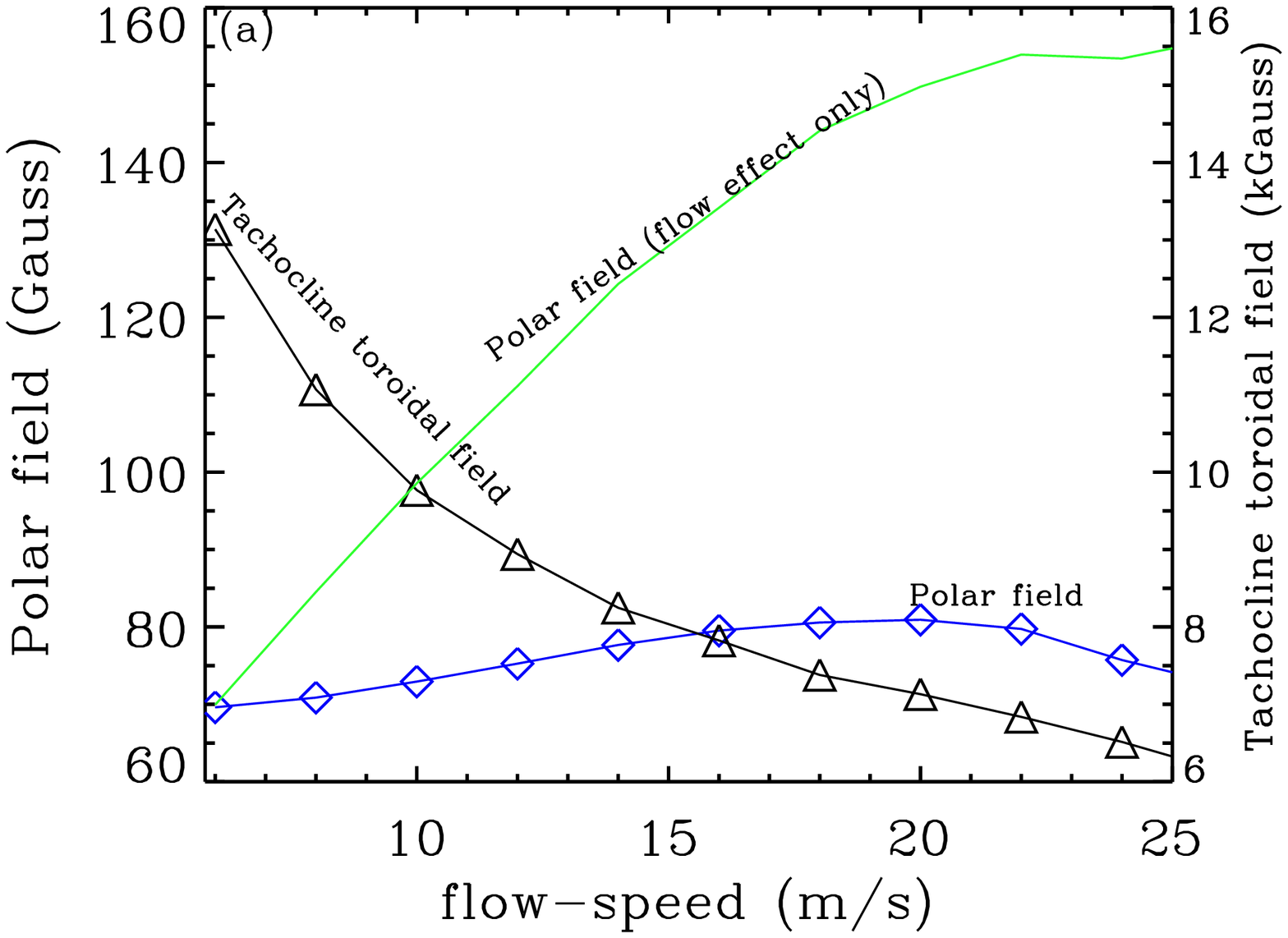}{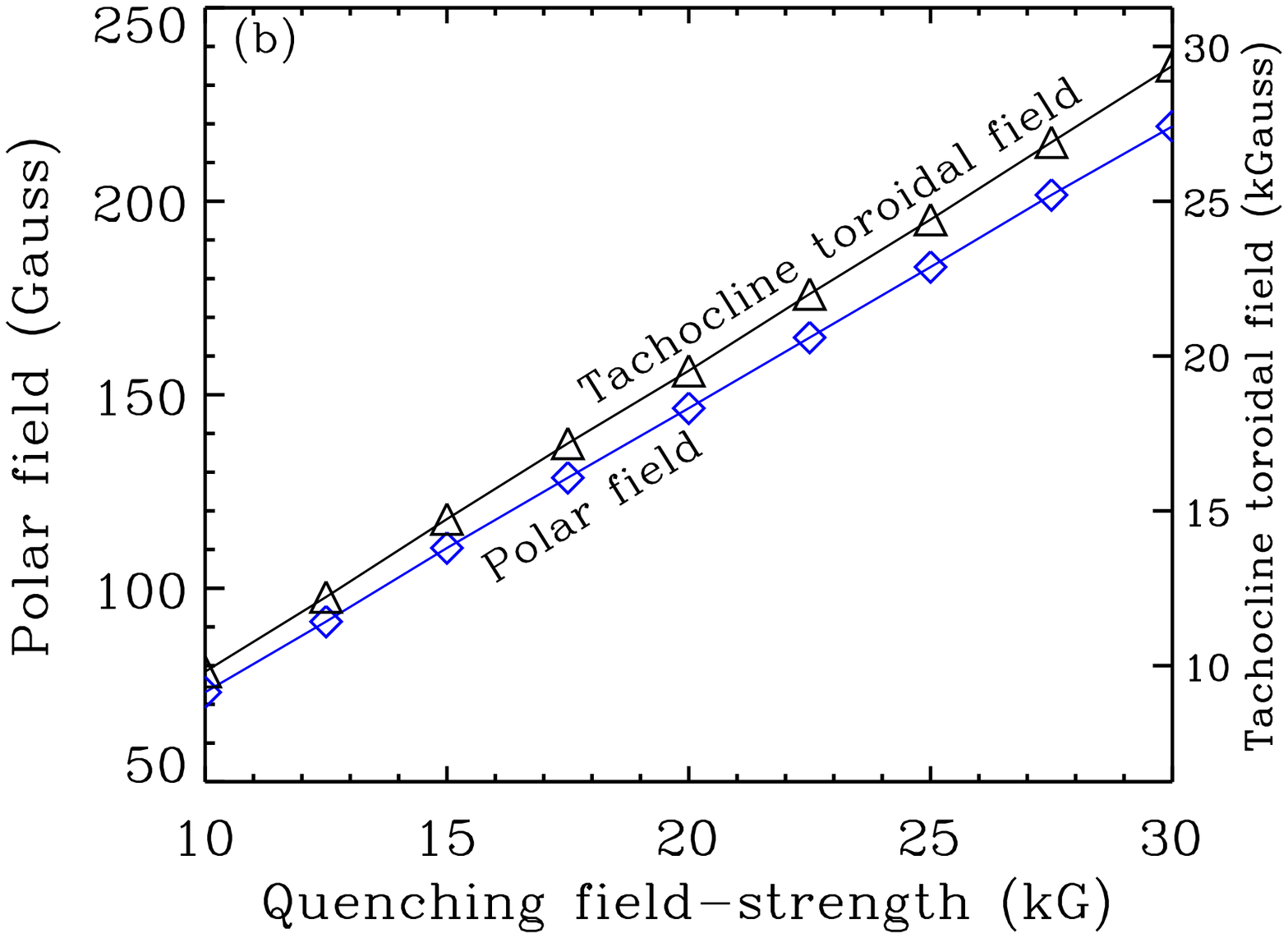}
\caption{(a) Polar fields at $\sim 75^{\circ}$ (left hand scale, blue 
diamonds and curve) and tachocline toroidal fields (right hand scale, 
black triangles and curve) for self-excited Babcock-Leighton flux transport 
dynamo solutions with different peak flow speeds (horizontal scale). Flow 
profile used is very similar to that taken in Dikpati \& Charbonneau (1999), 
which peaks at mid-latitudes. Green curve is polar fields normalized to the 
same tachocline toroidal field for all cases, illustrating polar fields 
obtained for the same tachocline toroidal fields.
(b) Polar fields at $\sim 75^{\circ}$ and sub-surface tachocline toroidal
fields as a function of quenching field strength.
}
\label{dynamo_polarfld}
\end{figure}

Since the dynamo-generated field strengths depend on the
choice of quenching field strength also, we plot in Figure 2b the
polar and subsurface toroidal field amplitudes as function of
the quenching field strength in the $\alpha$-effect. We find a nearly
linear relationship between the simulated magnetic field amplitudes
and quenching parameter. It is not yet possible to know from observations
what should be the actual quenching field strength, but to obtain
an optimum results of polar as well as subsurface toroidal field values
from a Babcock-Leighton flux-transport dynamo, it is conceivable that
the quenching field strengths range within 10-30 kGauss. Thus Figure 2(b)
implies that the polar and subsurface toroidal fields also be scaled
accordingly when a different quenching field strength is used.

We see from Figure 2a that for the example we have chosen, whether the
polar field is larger or smaller (blue diamonds and curve) for a larger
meridional flow depends on the amplitude of the flow. This effect was
also seen in \citet{bsss2004} who, using a surface transport model, 
reported non-monotonic behavior of the polar fields as a function of
flow-speed. This confirms that surface transport models and flux-transport 
dynamo models, despite operating with slightly different ingredients,
produce similar basic surface radial field features. 
For flux-transport dynamos, the amplitude of polar fields is affected 
greatly by the decrease in induced toroidal field at the bottom (black 
triangles and curve), because this reduces the surface poloidal source from 
which the polar fields are produced. If we normalize out this effect by
adjusting the bottom toroidal fields to the same value for all
meridional flow speeds, so the surface poloidal source is independent of
meridional flow speed, then the polar field would be given by the green
curve. Thus for the same surface source, the polar field does increase
markedly with an increase in meridional flow speed for the choice of
flow profile that peaks at mid-latitudes, consistent with the schematic
description provided in Figure 1. 

One important issue about a flux-transport dynamo driven by a 
Babcock-Leighton $\alpha$-effect is that it cannot really reproduce a best
match of polar field amplitudes with observations. This fact has been
noted by many Babcock-Leighton dynamo modelers since early 1990's
\citep{durney95, durney96, durney97, dc99}. In some attempts for
reproducing a correct polar field value using such models, the dynamo
has actually died after a few cycles, because a recycled polar flux
of only ten Gauss  is not enough to maintain the dynamo, for example
see the discussion of Table IV of \citet{dctg2002}. That consideration 
led to the development of a calibrated flux-transport dynamo driven by a 
small amount of the tachocline $\alpha$-effect in addition to a 
Babcock-Leighton $\alpha$-effect, results from which we discuss next.

\subsection{Simulations from a calibrated flux-transport dynamo
with sudden change in meridional flow-speed} 

We carry out further numerical simulations making a sudden change in 
meridional flow in the model and finding the change in the polar field 
in the first one or two cycles that follow. This is done for the case
of a self-excited flux-transport dynamo that achieves saturation only 
by 'alpha-quenching-like' amplitude-limiting nonlinearity, run with a
steady meridional flow and with the fixed dynamo source terms. In this
case, for a given set of input parameters, the model produces all
cycles of the same amplitude. 

Starting from such a saturated dynamo solution, we do numerical 
experiments for peak meridional circulation amplitudes ranging from 
$6-32 \,{\rm m}{\rm s}^{-1}$ for two distinct circulation profiles, one 
peaking at $25^{\circ}$ as used in Dikpati et al (2010a), and the
other peaking at $50^{\circ}$. Following expression (1) of Dikpati et al.
(2010a), we prescribe a meridional flow profile as:

$${\psi}r\sin\theta={\psi}_0(\theta-{\theta}_0)\sin\left[{\pi(r-R_b) \over
(R-R_b)}\right] \left(1-e^{-{\beta}_1 r{\theta}^{\epsilon}}\right)
\left(1-e^{{\beta}_2 r(\theta-\pi/2)}\right)
e^{-{((r-r_0)/\Gamma)}^2}, \quad\eqno(1)$$

\noindent in which, $R_b=0.69 R$, $\beta_1
\footnote{The parameter values for $\beta_1$, $\beta_2$ and $\Gamma$ 
were given in dimensionless units in the GRL paper by Dikpati et al. 
(2010a), whereas other parameters were given in dimensional units; we 
thank Dr. Luis Eduardo Antunes Vieira for helping us catching that.}
=0.1/(1.09 \times 10^{10}) \, {\rm cm}^{-1}$, 
$\beta_2=0.3/(1.09 \times 10^{10}) \,{\rm cm}^{-1}$, 
$\epsilon=2.00000001$, $r_0=(R-R_b)/5$, $\Gamma=3 \times 1.09 \times 
10^{10}\,{\rm cm}$ and $\theta_0=0$. This choice of the set of 
parameter values produce a flow pattern that peaks at $24^{\circ}$ as 
shown in the blue curve in Figure 3(a). The dimensionless length in our
calculation is $1.09 \times 10^{10} \,{\rm cm}$ and the dimensionless
time is $1.1\times 10^8 \,{\rm s}$, which respectively come from taking
the dynamo wavenumber, $k=9.2 \times 10^{-11} \, {\rm cm}^{-1}$, as one
dimensionless length, and the dynamo frequency, $\nu=9.1 \times 10^{-9}
\, {\rm s}^{-1}$, as one dimensionless time. In other words, the
dynamo wavelength ($2\pi \times 1.09 \times 10^{10} \, {\rm cm}$)
is $2\pi$ and the mean dynamo cycle period (22 years) is
$2\pi$ in our dimensionless units. Thus, in nondimensional units,
the above parameters are: $R_b=4.41$, $\beta_1=0.1$, $\beta_2=0.3$, 
$\epsilon=2.00000001$, $r_0=(R-R_b)/5$, $\Gamma=3$ and $\theta_0=0$. 
Thus in nondimensional units, by changing $\beta_1$ from 0.1 to 0.8 
and $\beta_2$ from 0.3 to 0.1, a flow pattern peaking at $50^{\circ}$ 
can be obtained. This means the dimensional values for $\beta_1$ and 
$\beta_2$ will be $\beta_1=0.8/(1.09 \times 10^{10}) \,{\rm cm}^{-1}$
and $\beta_2=0.1/(1.09 \times 10^{10}) \,{\rm cm}^{-1}$ in order to 
create a poleward surface flow peaking at $50^{\circ}$, as shown in
the red curve in Figure 3(a).

Here $\rho(r) = \rho_b {\left[ (R/r)-0.97 \right]}^m$, with $m=1.5$ 
and $\rho_b$ taken as 1 gm/cc for practical purposes. Then using
the constraint of mass-conservation, the velocity components can be
obtained as, $v_r={1\over \rho r^2 \sin\theta}{\partial \over \partial 
\theta} (\psi r \sin\theta)$ and $v_{\theta}= -{1 \over \rho r 
\sin\theta}{\partial \over \partial r} (\psi r \sin\theta).$ 

All other ingredients ($\alpha$-effects, differential rotation, and 
quenching) remain the same as in \citet{dgdu2010}. In both 
cases the peak velocity before the abrupt change is 18 m/s, which 
is set by adjusting the values of $\psi_0$. In both cases, $\theta_0=0$ 
ensures that there is a single meridional flow cell that extends all 
the way to the poles from the equator. In all cases, the change in 
meridional flow is timed to occur at the epoch of polar reversal. We 
get similar results if the change in flow speed occurs at other 
phases of a cycle. 

\begin{figure}[hbt]
\epsscale{1.0}
\plottwo{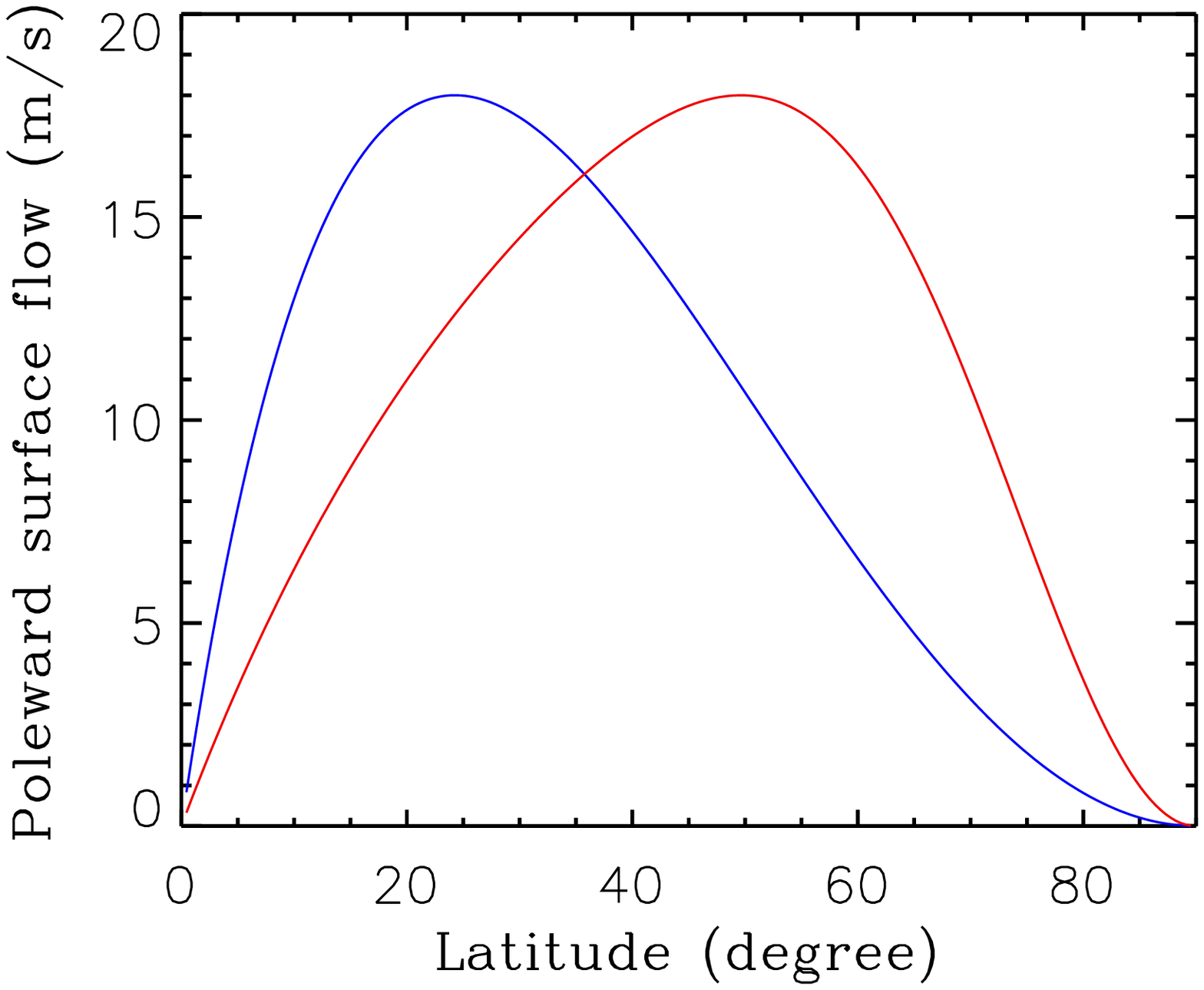}{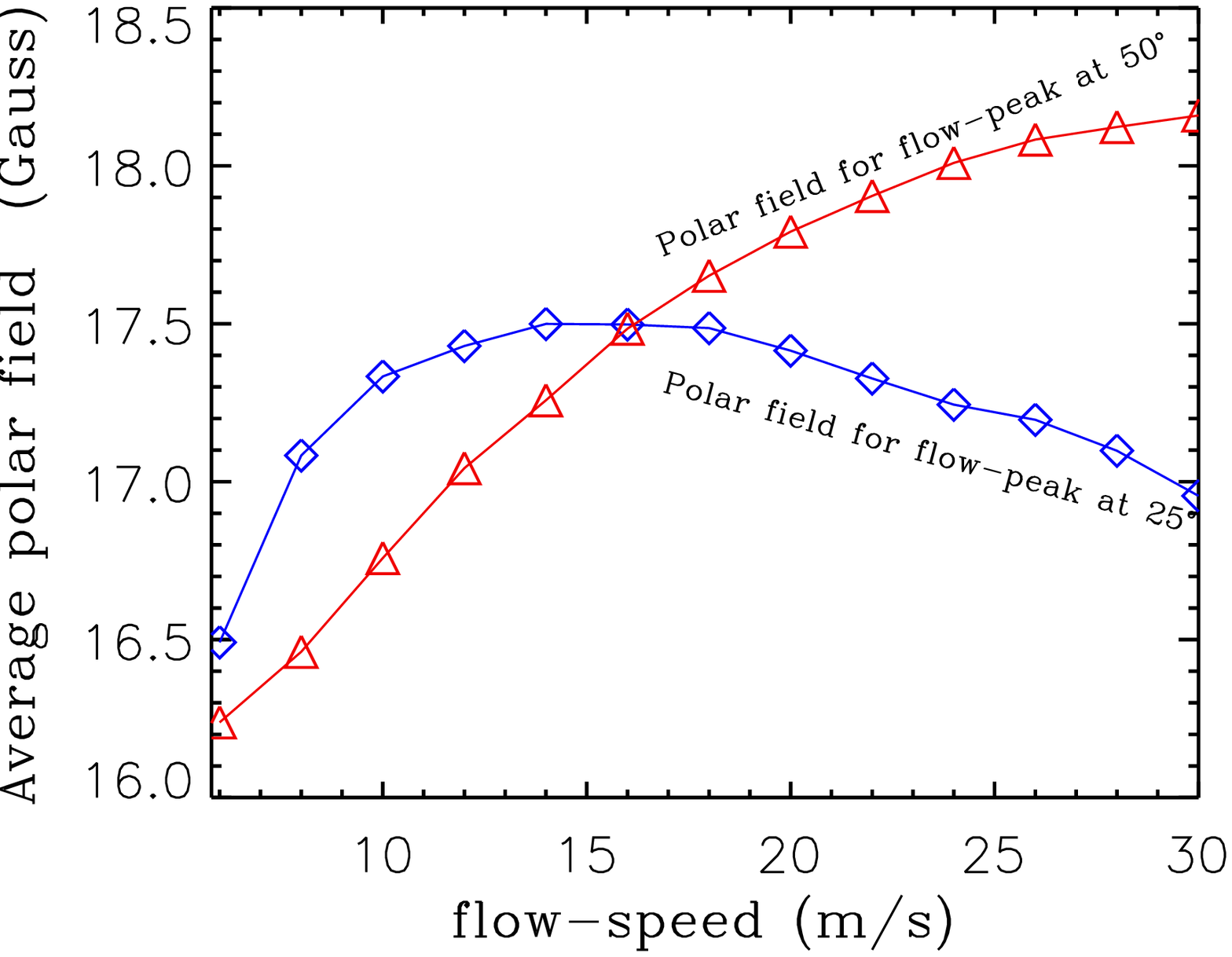}
\caption{(a) Typical profiles of poleward surface flow in meridional
circulation used in set of self-excited flux transport dynamo
simulations; blue curve from Dikpati et al (2010a) with peak near
$25^{\circ}$; red curve a profile with peak near $50^{\circ}$. (b)
Polar field amplitudes in the polar cap (latitudes $55^{\circ}$ to
$90^{\circ}$) from model simulations, for cycle immediately
following sudden change in meridional flow speed from 18 m/s to value
shown on horizontal axis. Nine Gauss has been added to the red curve
so it can be compared to the blue curve on the same scale. The much
lower actual polar fields in the red curve are probably caused by
the longer transport time from the source in active latitudes to
the poles, since the meridional flow peaks at a much higher latitude in
this case.
}
\label{polarfld_variation}
\end{figure}

The polar field amplitude produced by the model for all these cases is
plotted in Figure 3(b). This amplitude is the maximum, area weighted
average of the radial field from $55^{\circ}$ latitude to the pole for
the first polar field peak to occur after the change in meridional flow
speed (the same average as used in \citet{dgdu2010}). We see in
Figure 3(b) that the amplitude of the change in polar fields with a sudden
change in meridional flow is rather small -- no more than $10\%$ for the
full range of circulation amplitude change for either profile, perhaps
within the measurement error for average polar fields.  Thus the sudden
change in meridional flow has little effect on the polar field peak in
the cycle immediately following the change. Thus in a flux-transport 
dynamo model, a change in peak polar fields of as much as several tens 
of percent between one cycle and the next can not come from a change 
in meridional flow speed. It must come 
from a significant change in the amplitude of the source of polar
fields, namely the eruption and decay of active region magnetic flux.

\clearpage
\begin{figure}[hbt]
\epsscale{0.6}
\plotone{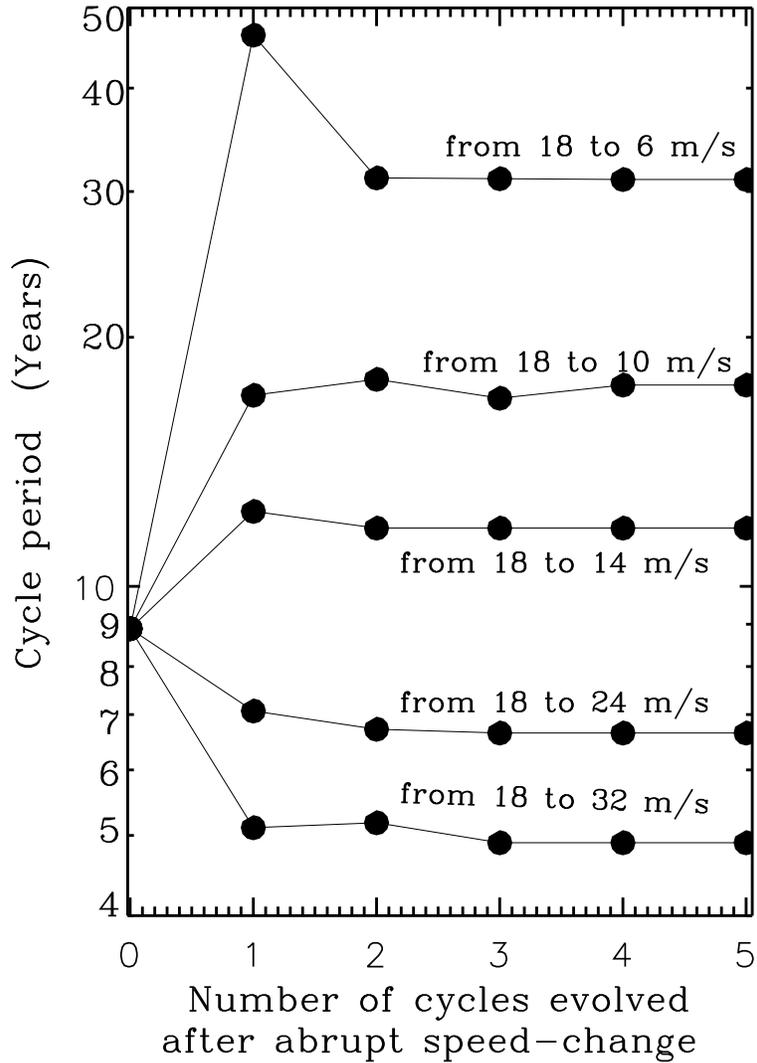}
\caption{Periods of first five cycles following a sudden change in 
meridional flow speed in a flux-transport dynamo model. The period 
sequence from each simulation is marked with the amount of the change. 
All simulations start from a previous simulation with peak meridional 
flow speed of $18\,{\rm m}{\rm s}^{-1}$, which gives a period of 
about 9 years, shown on the left hand scale for cycle number zero.}
\label{abrupt_cycle_change}
\end{figure}
\clearpage

How the polar field maximum of the next cycle changes with the change in
meridional flow is quite different for the two profiles. For the low
latitude peak in flow, both faster and slower meridional flow leads to a
decrease in polar field strength, while for the high latitude flow peak,
the faster(slower) the new flow, the larger (smaller) subsequent polar
field peak. The changes in polar field that occur for an increase in
meridional flow speed for the two meridional flow profiles are
completely consistent with the qualitative arguments made using
Figure 1 of \S2. So is the decline in polar field when the
meridional circulation with high latitude peak is reduced. Only the
decline in polar fields when the flow pattern peaking at low latitudes
is reduced requires a different explanation -- perhaps the dynamo cycle-period 
gets so long that there is more time for the surface flux moving to the poles
to be diffused down and not reach the pole.

These simulations of changes in polar fields due to drastic changes
in meridional circulation raise lead to an additional question -- 
how quickly does the dynamo period adjust to the changed meridional flow. 
Figure 4 plots the period of the first five cycles computed following 
the abrupt change in meridional flow speed without altering the
form of the streamlines. Not surprisingly, the new periods are about 
what we would expect for an advection-dominated flux-transport dynamo
in which the dynamo cycle-period is inversely proportional to the 
meridional flow-speed.  
What is perhaps surprising about the results seen in Figure 4 is how
quickly the model adjusts to the new period established by the
changed meridional flow. Except for the extreme case when the flow peak 
is reduced from $18\,{\rm m}{\rm s}^{-1}$ to  $6\,{\rm m}{\rm s}^{-1}$,
we see that the adjustment occurs almost entirely within the first cycle. 
A forthcoming paper on sequential data assimilation for solar dynamo 
models is addressing this issue in more detail; preliminary results
indicate that the 'response time' of a flux-transport dynamo to a 
change in meridional flow is as short as about 8 months.

The Figure 4 shows the settlement of the dynamo cycle-period for
a calibrated dynamo as discussed in\S3.2. For a pure
Babcock-Leighton flux-transport dynamo as in \citet{dc99}, without any
tachocline $\alpha$-effect, the cycle-period changes in a similar way. 
However, for the same meridional flow-speed the cycles are little
faster in that case.  
 
In diffusion-dominated dynamos the cycles are faster than that
in advection-dominated dynamos, due to enhanced diffusive transport
added to the advective transport of magnetic flux. An extensive 
analysis by \citet{hotta2010} shows how the dynamo cycle-period would 
change when the magnetic diffusivity in the bulk of the convection 
zone is increased. From the above study we anticipate that 
diffusion-dominated dynamos would respond to a sudden change in 
meridional flow-speed in an analogous way to that seen in Figure 3(b).
However, it would be worthwhile in the future to do an investigation 
of the response of advection-dominated dynamos of \citet{mnmy2010} that 
used a more sophisticated buoyancy mechanism than was used by \citet{dc99}.
  
\section{Discussion}

While both surface flux-transport models and flux-transport dynamo
models include the advective-diffusive transport of magnetic flux, and 
both can consistently explain polar field patterns, there
are inherent differences -- surface
transport models simulate the evolution of radial fields on the
photospheric latitude-longitude surface, whereas flux transport
dynamo models solve the axisymmetric dynamo equations for the toroidal
field, $B_T$, and the vector potential, $A$ ($\nabla \times A$ represents
poloidal fields) in the meridian plane in the convection
zone. There are additional physical effects operating in the evolution
of magnetic fields in a flux-transport dynamo model, due to the presence
of the radial flow in the circulation pattern, radial diffusion, and  
depth-dependent diffusivity. Similarly additional physical effects
are captured in surface transport models, namely their more realistic 
treatment of longitude dependence and hence estimation of
the Babcock-Leighton surface source term for poloidal and therefore
polar fields.

When poleward surface flow is increased in a flux-transport dynamo,
if there is no change in the profile of meridional flow, the upward flow
near the equator and the downward flow near the high-latitudes will also
increase, causing the poloidal fields produced from the leader polarity
to move up to a slightly higher diffusivity region compared to where
they would have been if the flow would not have increased. The poloidal
fields from the follower spots, on the other hand, do the opposite  ---
they sink down towards the lower diffusivity region (see Figure 1c). As
a consequence, the equatorward side of the polarity division line of
those poloidal fields undergoes faster diffusive decay, while the
poleward side undergoes less decay and therefore remains more frozen.
Since the dynamo equations solve for the vector potential, $A$, the
changes in $A$ described above get reflected in the radial component of
the poloidal field,  given by ${1 \over r^2 \sin\theta}{\partial
\over \partial\theta}\left(A\,r\,\sin\theta\right)$. Thus the presence
of depth-dependence in the diffusivity profile and the radial component
in the meridional flow profile both contribute to the increase in
the poloidal fields on the poleward side of the bipoles. This effect
is not present in surface transport models. 

Nevertheless, we have shown through physical description and numerical 
calculations that the surface flux-transport models and flux-transport 
dynamo models, despite some differences in their physics, produce 
similar results when run with a very similar latitudinal profile of 
surface poleward meridional flow, although surface-transport models
can better reproduce polar field patterns in latitude and time compared
to any dynamo model. 

As a consequence of the absence of depth-dependent diffusivity, the radial 
component of the meridional flow and the latitudinal component of the
poloidal fields, surface-transport models produce a more visible difference
in the polar field amplitudes than the polar fields obtained from a 
flux-transport dynamo model when a high-latitude reverse flow-cell is
present. \citet{jcss2009} have shown that the surface radial fields get 
significantly reduced at the pole if the poleward
surface flow reverses beyond $70^{\circ}$. However, flux-transport dynamo
models change the latitude location of the maximum surface radial fields
from pole to the boundary of the two cells when such models are run
with two flow cells \citep{bebr2005,jb2007,dgdu2010}. 
The observed polar fields from Wilcox Solar Observatory are instrumentally 
averaged over the latitude range from $55^{\circ}$ to the pole. Thus the 
model-derived radial fields integrated over the latitudes from $55^{\circ}$ 
up to the pole and weighted by the surface area, (i.e. $<B_r>={\int_0^{2\pi}
\int_{0}^{2\pi/5.143} B_r R^2 \sin\theta d\theta d\phi \over
\int_0^{2\pi} \int_{0}^{2\pi/5.143} R^2 \sin\theta d\theta d\phi}$)
are not actually much different in the two cases whether a high-latitude 
reverse-cell is present or absent (see the discussion in \citep{dgdu2010}). 

In this analysis, we confined ourselves to consideration of 
advection-dominated flux-transport dynamos and surface transport models. 
How polar fields respond to changes in meridional flow in the cases of
diffusion-dominated dynamos \citep{ynm2008} and flux-transport dynamos
with more complexities included, such as turbulent pumping \citep{gd2008},
diffusivity quenching \citep{gdd2009} and buoyancy-induced delay in surface
poloidal field generation \citep{jpl2010}, have not yet been explored.

An alternative explanation for the weak polar fields and long minimum of 
cycle 23 has been given by \citet{nmm2011}. Using an advection-dominated
flux-transport dynamo \citep{mnm2009} that operates with a two-step 
diffusivity profile, \citet{nmm2011} 
do a large number of simulations in which the meridional flow amplitude 
is changed randomly, once per cycle, at cycle maximum. The peak amplitude 
falls in the range of $15 - 30 \,{\rm m}{\rm s}^{-1}$. They find that 
cycles in which the meridional flow speed is larger in the ascending phase 
than in the declining phase tend to be followed by longer, deeper minima, 
and the polar field strength of such cycles tend to be weaker. Both these 
correlations are modest, with correlation coefficient $r \sim 0.5$, which
corresponds to a variance of one-quarter only. This means that the other
75\% of the variance in polar fields' weakening must be due to other effects.
In addition, no observational evidence was given in \citet{nmm2011} that 
the meridional circulation in cycle 23 actually was higher in the ascending 
phase than in the descending phase, and, indeed, the best measures of surface 
plasma flow for cycle 23 that exists, namely that of \citet{ulrich2010} and
\citet{ba2010}, does not support the assumption of speed-up in meridional 
flow-speed in the ascending phase of cycle 23 (see figure 6 of 
\citet{ulrich2010} and figure 3 of \citet{ba2010}). Thus it follows that, 
at least for cycle 23, it is not possible to explain the polar field drop 
or the long minimum using the correlation found by \citet{nmm2011}.

Currently all benchmarked flux-transport dynamos operate in the 2D 
axisymmetric regime (see \citet{jouveetal2008}). It is also necessary 
to investigate the role of longitude-dependence from the tilted, bipolar 
active regions in the generation of evolution of the Sun's polar fields 
using a 3D version of flux-transport dynamos.

\section{Acknowledgements:}

We thank Peter Gilman for reviewing the manuscript. We extend our thanks
to an anonymous referee for helpful comments which helped improve
the manuscript. This work is partially supported by NASA's Living With a Star
program through the grant NNX08AQ34G. The National Center for
Atmospheric Research is sponsored by the National Science Foundation.

\end{document}